\documentclass[journal=jpccck,manuscript=article]{achemso}
\usepackage{amsmath,color}
\usepackage{graphicx}
\usepackage{amssymb}
\usepackage{caption}
\usepackage{subcaption}
\usepackage{rotating}

\author{Albert J. Pool}
\affiliation[Institute for Theoretical Physics,Universiteit Utrecht]
{Institute for Theoretical Physics, Universiteit Utrecht,
Princetonplein 5, 3584 CC Utrecht, The Netherlands}

\author{Sandeep K. Jain}
\affiliation[Institute for Theoretical Physics,Universiteit Utrecht]
{Institute for Theoretical Physics, Universiteit Utrecht,
Princetonplein 5, 3584 CC Utrecht, The Netherlands}
\email{S.K.Jain@uu.nl}

\author{Gerard T. Barkema}
\affiliation[Department of Information and Computing Science, Universiteit Utrecht]
{Department of Information and Computing Science, Universiteit Utrecht,
Princetonplein 5, 3584 CC Utrecht, The Netherlands}
\email{G.T.Barkema@uu.nl}

\title{Structural characterization of carbon nanotubes via the vibrational density of states}

\begin{document}

\begin{abstract}
The electrical and chemical properties of carbon nanotubes vary significantly with different chirality and diameter, making the experimental determination of these structural properties important. Here, we show that the vibrational density of states (VDOS) contains information on the structure of carbon nanotubes, particularly at low frequencies. We show that the diameter and chirality of the nanotubes can be determined from the characteristic low frequency $L$ and $L'$ modes in the VDOS. For zigzag nanotubes, the $L$ peak splits into two peaks giving rise to another low energy $L''$ peak. The significant changes in the frequencies and relative intensities of these peaks open up a route to distinguish among structurally different nanotubes. A close study of different orientations of Stone--Wales defects with varying defect density reveals that different structural defects also leave distinct fingerprints in the VDOS, particularly in the $L$ and $L'$ modes. With our results, more structural information can be obtained from experiments which can directly measure the VDOS, such as inelastic electron and inelastic neutron spectroscopy.
\end{abstract}

\section{INTRODUCTION}
Carbon nanotubes (CNTs) have emerged as one of the most promising materials because of their peculiar and remarkable electronic, thermal and mechanical properties \cite{Saito1998}. Many applications are foreseen for carbon nanotubes in future nanotechnology and semiconductor industry, such as smaller transistors \cite{franklin2012sub,desai2016mos2}, and improved batteries \cite{lee2010high}; both of which are currently limiting factors in the development of better electronic devices. Recently, the use of carbon nanotubes has been shown in flexible integrated circuits \cite{sun2011flexible,wang2014tuning} and as promising candidates for conductors in nanoelectronics \cite{cui2009joining}. Another recent discovery is the use of carbon nanotubes as a cavity for ice, to increase its melting point by as much as $100^{0}$ C, but this effect critically depends on the diameter of the tube \cite{agrawal2016observation}. Distinct carbon nanotubes have shown very different electronic and thermal behavior \cite{wilder1998electronic,odom1998atomic}, therefore it is fundamentally and experimentally important to distinguish different types of tubes based on their chirality and diameter. 

Carbon nanotubes are characterized by two integer numbers $(n,m)$, describing the chiral vector between two equivalent points on a graphene sheet, which coincide when the sheet is rolled into a tube. Two important chiralities are armchair $(n,n)$, featuring C--C bonds perpendicular to the tube axis, and zigzag $(n,0)$, featuring C--C bonds parallel to the tube axis. A nanotube with any other $(n,m)$ chirality is called a chiral tube. \cite{dresselhaus1996science}

It is possible in experiments to identify a single carbon nanotube by its diameter and chiral angle as measured by direct local techniques such as atomic resolution scanning tunneling microscopy (STM) \cite{wilder1998electronic}, transmission electron microscopy (TEM) \cite{hashimoto2004direct,zhu2008atom}, and electron diffraction \cite{Iijima1993,zuo2003atomic,qin2007determination}. Other indirect and rather cheap methods to characterize carbon nanotubes are Raman spectroscopy \cite{jorio2001structural}, and photoluminescence excitation spectroscopy \cite{bachilo2002structure,lefebvre2007polarized,tsyboulski2007structure}. Very recently, the determination of chiral indices of two- or three-walled carbon nanotubes by a combination of electron diffraction and Raman spectroscopy has been reported \cite{levshov2017accurate}.

In experiments, intrinsic defects commonly occur in carbon nanotubes, such as the Stone--Wales (SW) defect which is formed by the rotation of a C--C bond by $90^{0}$ \cite{miyamoto2004spectroscopic,kotakoski2011stone}. This rotation results in two pentagons and two heptagons in the otherwise hexagonal lattice. This is a common defect in graphene as well \cite{araujo2012defects} and the formation energy of an SW defect in carbon nanotubes is lower than it is in graphene, and further decreases as a function of decreasing tube diameter \cite{kabir2016kinetics}. SW defects have direct consequences on the electrical \cite{song2015electronic}, mechanical \cite{lu2005effect,tserpes2007effect,sharma2014effect}, and chemical\cite{bettinger2005reactivity,robinson2006role} properties of carbon nanotubes. The SW defect can be considered to form spontaneously as a result of excess strain \cite{nardelli1998mechanism,jiang2004defect}, and is suggested to be responsible for plastic deformation of carbon nanotubes \cite{suenaga2007imaging}. This defect is considered as a dislocation dipole: rotation of further C--C bonds separates it into two dislocation cores both consisting of a pentagon and a heptagon \cite{samsonidze2002energetics}. Other nanotubes with topological defects, for instance containing octagons, have also been reported \cite{nardelli1998mechanism}.

Other types of defects in carbon nanotubes include vacancies, which show a stronger effect on the strength of a carbon nanotube \cite{sammalkorpi2005irradiation}. Vacancy defects are commonly formed by irradiation and can survive for macroscopic times at room temperature; but as a vacancy is metastable, atoms will rearrange to mend the defect \cite{krasheninnikov2001formation,charlier2002defects}. Carbon nanotubes can also contain atoms other than carbon; common hetero-atom dopants experimentally integrated into the graphene lattice include nitrogen and boron. Due to the different electrical and chemical properties of doped carbon nanotubes, applications are suggested in chemistry and energy storage devices \cite{zhang2014substitutional}. It is also possible for foreign atoms or molecules to be adsorbed to the carbon nanotube surface; for many molecules adsorption occurs most strongly at defect sites \cite{robinson2006role,kim2007defect}. This effect is studied to use carbon nanotubes as chemical sensors \cite{robinson2006role,mittal2014carbon}.

Computer simulation is a powerful tool in carbon nanotube research. By simulations, the properties can be predicted of carbon nanotubes of any chirality or any number of defects: many different nanotubes can be studied without having to synthesize every single one of them. A common and frequently used method to simulate carbon nanotubes is density functional theory (DFT). However, DFT is computationally intensive and therefore limits the size of the system that can be studied \cite{cerqueira2014density,ayissi2015preferential}. A semiempirical potential allows to study much larger systems. Small systems of nanotubes have particularly poorly resolved low frequency vibrational spectra therefore to resolve this regime of VDOS nicely, large system sizes are required. In the case of vibrational spectrum studies; it is particularly important because a system of $N$ atoms has $3N$ vibrational degrees of freedom\cite{saito1998raman,rols2012unravelling}, therefore the vibrational spectrum of a small system is only a crude approximation of the spectrum of a macroscopically large carbon nanotube. Semiempirical methods have recently been shown to correspond fairly well with experiments \cite{do2015calculation,tailor2017empirical}.

The vibrational density of states (VDOS) can well be used to determine many chemical and physical properties of carbon nanotubes. Besides studying chirality and defects, the VDOS can also be used to calculate thermal conductivity \cite{yao2005thermal}. Many properties of carbon nanotubes have been studied using Raman spectroscopy such as chirality and defects \cite{jorio2001structural,sauvajol2002phonons,miyamoto2004spectroscopic,saidi2014probing,herziger2014double,laudenbach2016diameter}, but with this technique only certain Raman active phonon modes can be observed \cite{rao1997diameter,saito1998raman,sauvajol2002phonons}. It is possible to study the full vibrational spectrum in experiments by inelastic neutron scattering \cite{rols2000phonon,sauvajol2002phonons}, or inelastic electron tunneling spectroscopy (IETS) \cite{kislitsyn2014vibrational}. IETS can also be used for localized measurements using the tip of an STM (STM-IETS), which allows the study of local properties such as defects, different tubes joined together, or tube caps \cite{vitali2004phonon,vandescuren2007characterization}. The vibrational spectrum of graphene has already been experimentally mapped with IETS \cite{natterer2015strong}, and this opens up a possibility to use the same technique to characterize carbon nanotubes.

In this paper, we show that a semiempirical potential developed for graphene \cite{jain2015strong} can be effectively used for carbon nanotubes. The vibrational spectrum of a carbon nanotube as calculated by the simple elastic potential; approaches that of graphene for large diameters. We show that the armchair and zigzag nanotubes have distinct low frequency vibrational modes and based on these modes one can characterize the chirality of a carbon nanotube. These low frequency phonon modes are very sensitive to the tube diameters and therefore can be used to determine the diameter of a nanotube. We further show that the density of the point defects such as Stone--Wales defects can be measured by simple study of relative intensities of low energy peaks in vibrational spectrum.

\section{METHOD}
To calculate the vibrational spectrum of carbon nanotubes, we use a recently developed semiempirical potential for graphene \cite{jain2015strong} given by
\begin{equation}
E=\frac{3}{16}\frac{\alpha}{d^2}\sum_{i,j}(r_{ij}^2-d^2)^2+\frac{3}{8}\beta d^2\sum_{j,i,k}\left(\theta_{jik}-\frac{2\pi}{3}\right)^2+\gamma\sum_{i,jkl}r_{i,jlk}^2.
\end{equation}
Here, $r_{ij}$ is the length of the bond between two atoms $i$ and $j$, $\theta_{jik}$ the angle between the two bonds connecting atom $i$ to $j$ and $k$, respectively, and $r_{i,jkl}$ the distance between atom $i$ and the plane formed by its three neighboring atoms $j$, $k$, and $l$. The parameters $\alpha=26.060\text{ eV/\AA}^2$, $\beta=5.511\text{ eV/\AA}^2$, $\gamma=0.517\text{ eV/\AA}^2$, and $d=1.420\text{ \AA}$ are used, which were obtained by density-functional theory (DFT) calculations for graphene \cite{jain2015strong}. This potential was effectively used to study twisted bilayer graphene \cite{Jain2016}, vibrational properties of graphene \cite{jain2015probing}, effect of boundary conditions in graphene samples \cite{jain2016boundaries}, and graphene nanobubbles \cite{Jain2016a}.

Using this potential the Hessian matrix is obtained containing the second derivatives of the energy with respect to the $x$-, $y$- and $z$-coordinates of all atoms. The eigenvalues of this matrix represent the force constants of the vibrational modes in $\text{eV/\AA}^2$, and the eigenvectors represent the normal coordinates of the vibrational modes. These force constants $k$ are converted into frequencies as $f=\frac{1}{2\pi c} \sqrt{\frac{k}{m}}$, with $c$ the speed of light and $m=1.99\times10^{-26}\text{ kg}$, the mass of a carbon atom. In our simulations we use force-free periodic boundary conditions \cite{jain2016boundaries} in the direction along the tube. There are three translational modes and one rotational mode around the axis, assigned by their force constants which are zero. The area under the VDOS plot is equal to the total number of the vibrational modes.

Using this method, the VDOS is obtained for many carbon nanotubes having 600 atoms with different chiralities and defect densities. The VDOS is convoluted with a Gaussian function with a width $\sigma=14\text{ cm}^{-1}$, a value chosen for comparison with earlier results for graphene \cite{jain2015probing}. 

\section{RESULTS AND DISCUSSION}
We calculate the VDOS for many carbon nanotubes of different diameters and chiral angles. In particular, we show the dependency of VDOS on the diameter of armchair and zigzag nanotubes. We also show some soft vibrational modes of the tubes which can be used to characterize carbon nanotubes. The vibrational spectrum for different chiral angles has a unique signature in the low frequency vibrational modes. Furthermore, we show that the relative intensities of low frequency vibrational peaks can be used to determine the Stone--Wales defect densities for armchair and zigzag nanotubes.

\subsection{Armchair nanotubes}
\begin{figure*}
\begin{subfigure}{0.6\textwidth}
(a)\\
\includegraphics[scale=1]{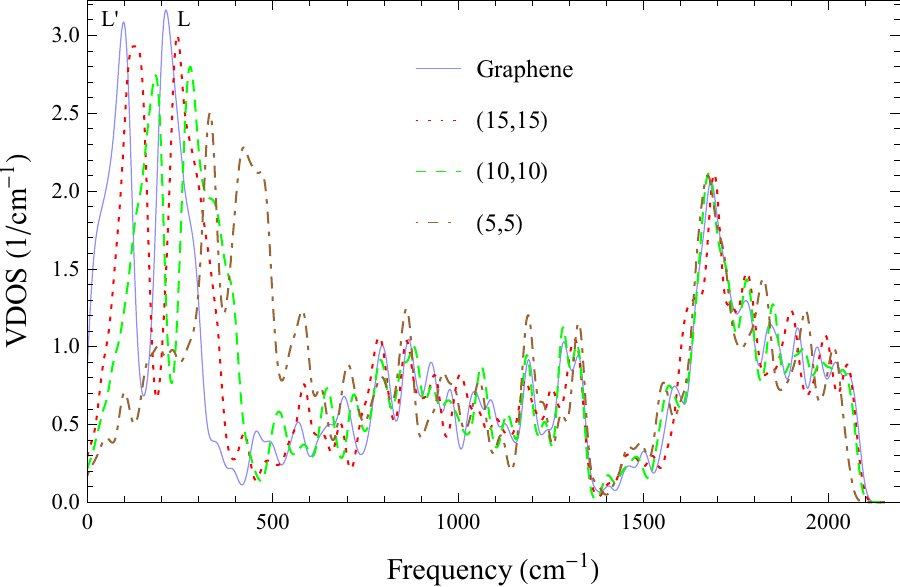}
\end{subfigure}
~
\begin{minipage}{0.36\textwidth}
(b)
\begin{subfigure}{.89\textwidth}
\includegraphics[scale=1]{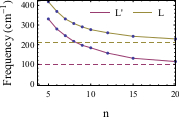}
\end{subfigure}\\
\vspace{10pt}\\
(c)
\begin{subfigure}{.9\textwidth}
\includegraphics[scale=1]{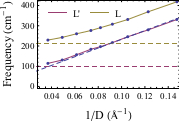}
\end{subfigure}
\end{minipage}

\begin{center}
(d)
\begin{subfigure}{0.25\textwidth}
\includegraphics[scale=.4,trim=10cm 5cm 9cm 6cm,clip=true]{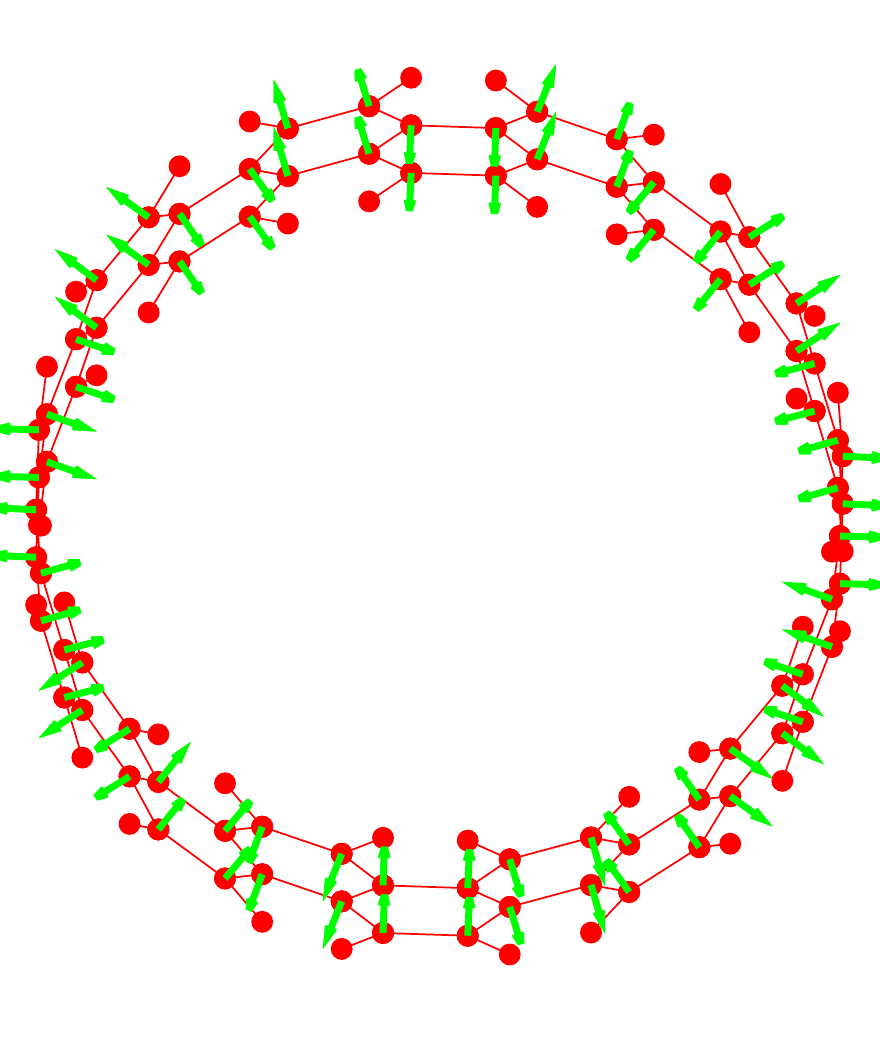}
\end{subfigure}
~
(e)
\begin{subfigure}{0.25\textwidth}
\includegraphics[scale=.4,trim=10cm 5cm 9cm 6cm,clip=true]{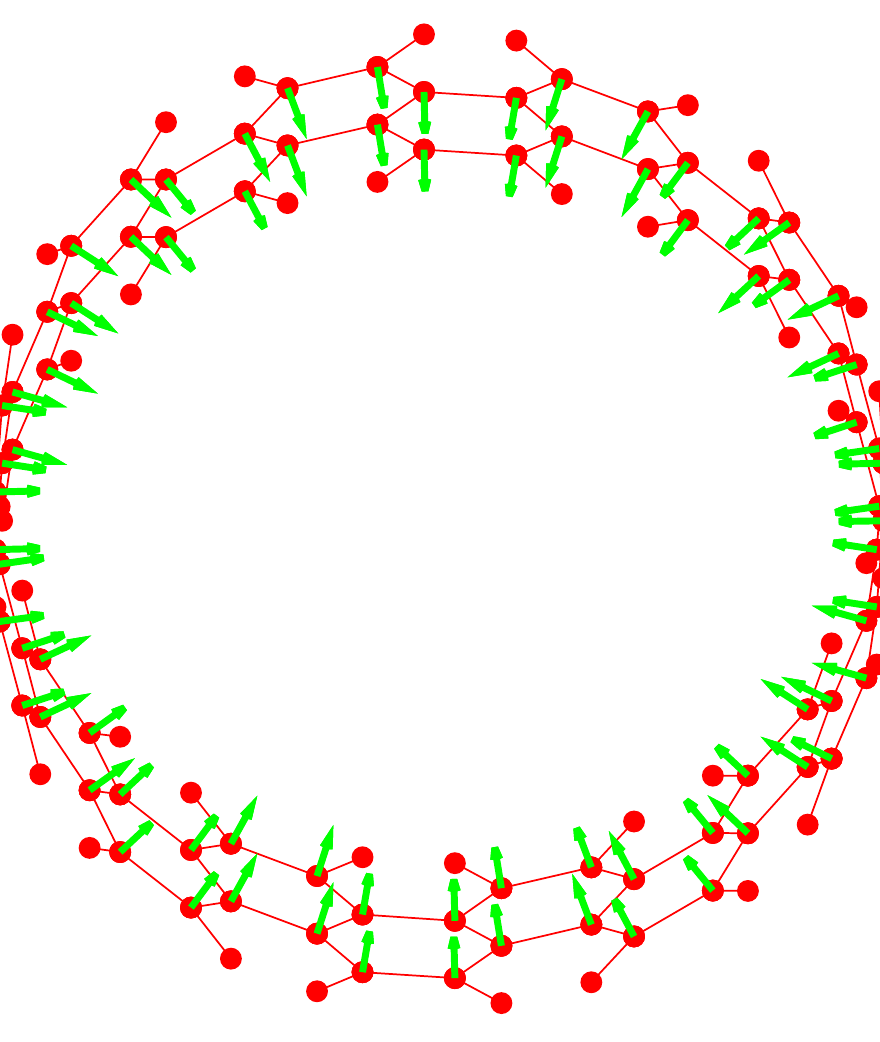}
\end{subfigure}
\end{center}
\caption{Dependence of VDOS for armchair carbon nanotubes on the diameter $D$. (a) VDOS of three armchair nanotubes compared to graphene. (b) Position of $L$ and $L'$ peaks as a function of chiral index $n$ for nanotubes with chirality $(n,n)$ . Dashed lines are for graphene. (c) Position of $L$ and $L'$ peaks as a function of inverse tube diameter, with the radial breathing mode frequency from eq.~\eqref{eq:rbm} shown in dashed blue. (d) One of the vibrational modes in the $L$ peak of a $(10,10)$ nanotube. The mode corresponds to the rotative motion of the axial bonds out of the carbon plane. Here, red dots are carbon atoms in the nanotube and green arrows are corresponding vibrational displacements. (e) The radial breathing mode in the $L'$ peak of a $(10,10)$ nanotube.}
\label{fig:armchair}
\end{figure*}

We calculate the vibrational spectrum of various armchair nanotubes of different diameters ($D$) and compare them with the VDOS of graphene by Jain et al. \cite{jain2015probing} as shown in figure~\ref{fig:armchair}. We show that the small diameter tube $(5,5)$ has high frequencies for the $L$ and $L'$ modes and there is redshift in the peaks as the diameter of the tubes increases and converges to the values of graphene in infinite diameter limit. The VDOS of three armchair nanotubes and graphene has been plotted as shown in figure~\ref{fig:armchair}(a). The frequencies of the $L$ and $L'$ peaks for many armchair tubes as a function of $(n,n)$ chiral index, and as a function of inverse tube diameter are shown in figure~\ref{fig:armchair}(b) and figure~\ref{fig:armchair}(c), respectively. For figure~\ref{fig:armchair}(c), the diameter $D$ of an $(n,m)$ tube is computed using
\begin{equation}\label{eq:diameter}
D=\frac{\sqrt{3}}{\pi}d\sqrt{n^2+nm+m^2},
\end{equation}
with $d=1.420$ \AA, the ideal C--C bond length \cite{dresselhaus1996science}. 

At high frequencies, the VDOS is very similar to that of graphene but the variation occurs in the low frequency range ($0-500~ cm^{-1}$). It is apparent that the $L$ and $L'$ peaks shift towards a lower frequency as a function of increasing diameter. The frequency of the $L$ peak for the smallest diameter $(5,5)$ tube ($D=6.78$ \AA) is around $99\%$ higher compared to that of graphene but rapidly decreases to $31\%$ for the $(10,10)$ tube ($D=13.56$ \AA) and $15\%$ for the $(15,15)$ tube ($D=20.34$ \AA). The variation in the frequency of the $L'$ peak is even more prominent compared to graphene as the values are: $240\%$ for the $(5,5)$ tube, $89\%$ for the $(10,10)$ tube and $34\%$ for the $(15,15)$ tube. There is a significant decrease in the intensities of both $L$ and $L'$ peaks as a function of decreasing diameter. The intensity of the $L$ peak decreases by $5\%$ for the $(15,15)$ tube, $12\%$ for the $(10,10)$ tube, and $28\%$ for the $(5,5)$ tube, compared to bulk graphene. Similarly, the intensity of $L'$ peak decreases by $5\%$ for the $(15,15)$ tube, $11\%$ for the $(10,10)$ tube, and $19\%$ for the $(5,5)$ tube, compared to bulk graphene.         

To further characterize these two low energy vibrational modes, we show vibrational modes found in the $L$ and $L'$ peaks as shown in figure~\ref{fig:armchair}(d) and figure~\ref{fig:armchair}(e), respectively. One of the vibrational modes in the $L'$ peak is the radial breathing mode (RBM), a Raman active mode for which the frequency has been determined experimentally for many carbon nanotubes \cite{sauvajol2002phonons,maultzsch2005radial}. 
The radial breathing mode frequency is given by Maultzsch et al.\cite{maultzsch2005radial} as
\begin{equation}\label{eq:rbm}
\omega_{RBM}=\frac{c_1}{D}+c_2,
\end{equation}
with experimentally determined constants $c_1=215\pm2\text{ cm}^{-1}\text{nm}$ and $c_2=18\pm2\text{ cm}^{-1}$, and $D$ is determined by eq.~\eqref{eq:diameter}. It is interesting to note that the position of the $L'$ peak in armchair nanotubes complies almost exactly with this experimental radial breathing mode frequency, as seen in figure~\ref{fig:armchair}(c). This observation further validates our numerical results for carbon nanotubes.

\subsection{Zigzag nanotubes}
\begin{figure*}
\begin{subfigure}{0.6\textwidth}
(a)\\
\includegraphics[scale=1]{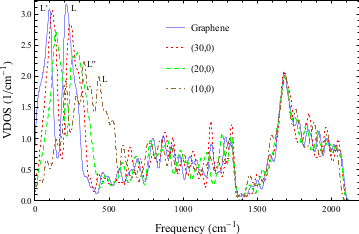}
\end{subfigure}
~
\begin{minipage}{0.36\textwidth}
(b)
\begin{subfigure}{.89\textwidth}
\includegraphics[scale=1]{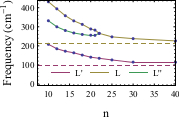}
\end{subfigure}\\
\vspace{10pt}\\
(c)
\begin{subfigure}{.9\textwidth}
\includegraphics[scale=1]{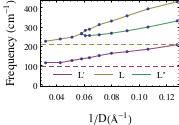}
\end{subfigure}
\end{minipage}

\begin{center}
(d)
\begin{subfigure}{0.25\textwidth}
\includegraphics[scale=.4,trim=10cm 5cm 9cm 9cm,clip=true]{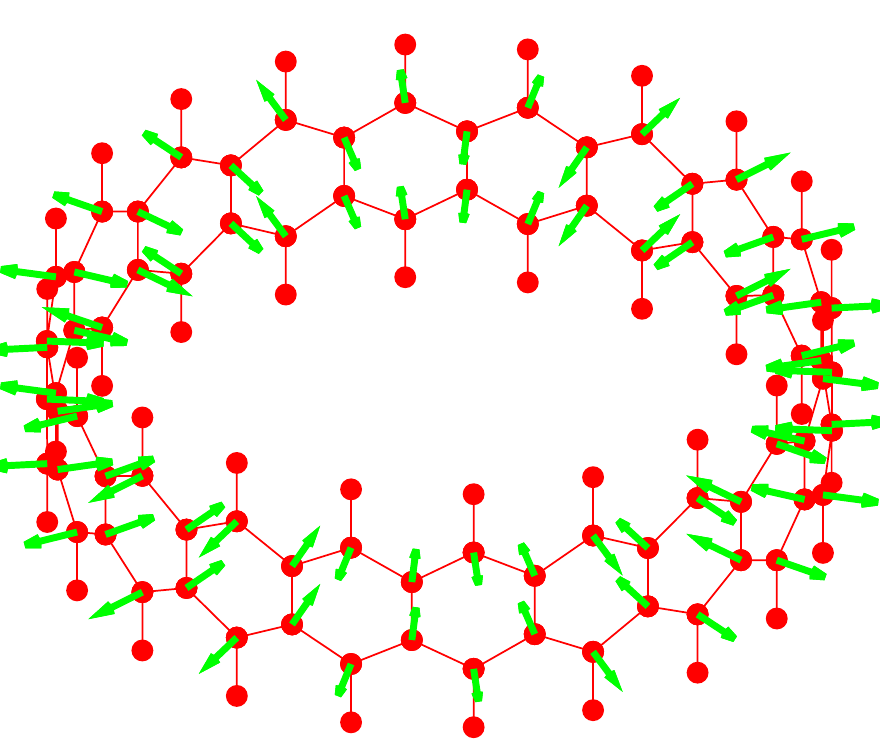}
\end{subfigure}
~
(e)
\begin{subfigure}{0.25\textwidth}
\includegraphics[scale=.4,trim=98mm 48mm 89mm 8cm,clip=true]{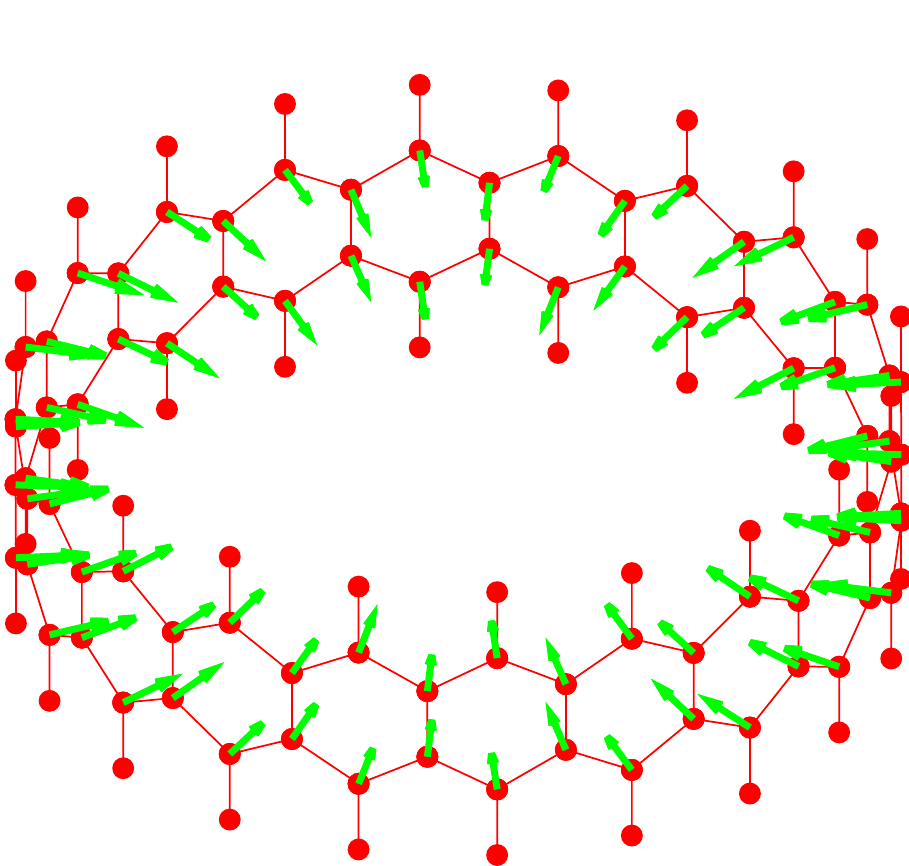}
\end{subfigure}
~
(f)
\begin{subfigure}{0.25\textwidth}
\includegraphics[scale=.4,trim=10cm 5cm 9cm 9cm,clip=true]{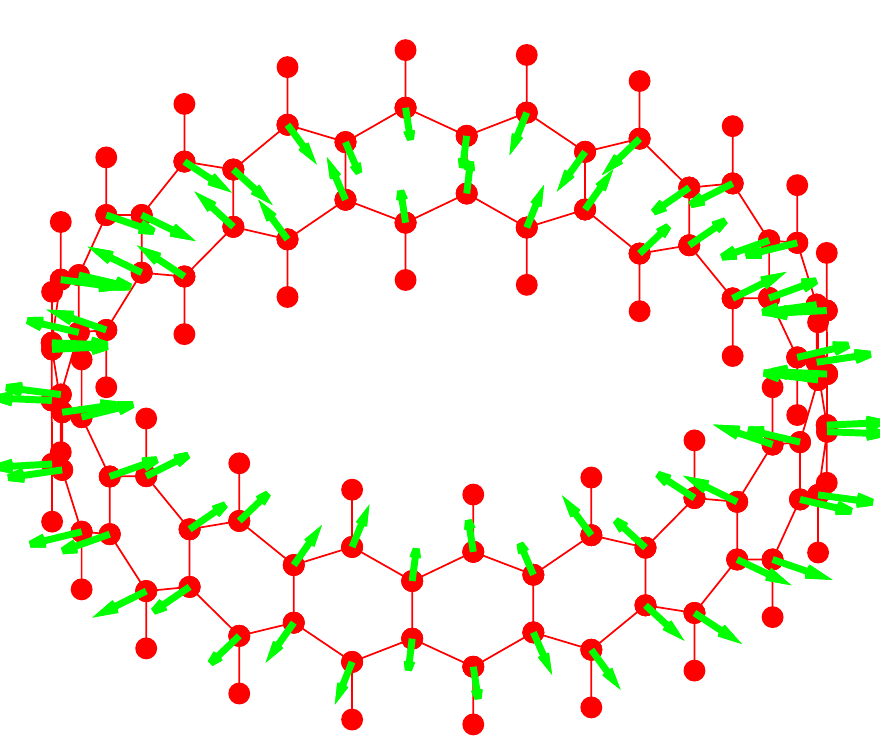}
\end{subfigure}
\end{center}
\caption{Dependence of VDOS for zigzag carbon nanotubes on the diameter $D$.(a) VDOS of three zigzag nanotubes compared to graphene. (b) Position of $L$, $L'$ and $L''$ peaks as a function of chiral index $n$ for nanotubes with chirality $(n,0)$. Dashed lines are for graphene. (c) Position of $L$, $L'$ and $L''$ peaks as a function of inverse tube diameter. (d) One of the vibrational modes found in the $L$ peak of a $(20,0)$ nanotube. The mode corresponds to the rotative motion of the circumferential bonds out of the carbon plane. Here, red dots are carbon atoms in the nanotube and green arrows are corresponding vibrational displacements. (e) The radial breathing mode in the $L'$ peak of a $(20,0)$ nanotube. (f) One of the vibrational modes found in the $L''$ peak of a $(20,0)$ nanotube. The mode corresponds to the rotative motion of the axial bonds out of the carbon plane.}
\label{fig:zigzag}
\end{figure*}

We further calculate the vibrational spectrum of various zigzag nanotubes of different dia\-meters and compare them with the VDOS of graphene as shown in figure~\ref{fig:zigzag}. Interestingly we find that there are three, rather than two distinct low frequency peaks visible for $n\lesssim20$ tubes. In these small diameter zigzag tubes ($D\lesssim 15$ \AA), the $L$ peak splits into two: we label the one with the lower frequency as $L''$. We show that the small diameter tube $(10,0)$ has high frequencies for the $L$ and $L'$ modes and there is a redshift in the peaks as the diameter of the tubes increases; it converges to the values of graphene in the infinite diameter limit. The VDOS of three zigzag nanotubes and graphene has been plotted in figure~\ref{fig:zigzag}(a). Figure~\ref{fig:zigzag}(b) and figure~\ref{fig:zigzag}(c) show the frequencies of the $L$, $L'$, and $L''$ peaks for many zigzag tubes as a function of $(n,0)$ chiral index, and as a function of inverse tube diameter, respectively.

At high frequencies, the VDOS is very similar to that of graphene but the variation occurs in the low frequency range ($0-600~ cm^{-1}$). We show that the $L$ and $L'$ peaks shift towards a lower frequency as a function of increasing diameter. The frequency of the $L$ peak for the smallest diameter $(10,0)$ tube ($D=7.83$ \AA) is around $104\%$ higher compared to that of graphene but rapidly decreases to $12\%$ for the $(30,0)$ tube ($D=23.48$ \AA). The intensity of this peak decreases by $10\%$ for the $(30,0)$ tube, to $36\%$ for the $(10,0)$ tube compared to graphene. The variation in the frequency of the $L'$ peak is even more prominent as the values are: $114\%$ for the $(10,0)$ tube, $44\%$ for the $(20,0)$ tube and $18\%$ for the $(30,0)$ tube. There is a significant decrease in the intensities of the $L'$ peak as a function of decreasing diameter. The intensity of the $L'$ peak decreases by $2\%$ for the $(30,0)$ tube, $12\%$ for the $(20,0)$ tube, and $37\%$ for the $(10,0)$, compared to bulk graphene.  

The mode shown in figure~\ref{fig:zigzag}(f) is a vibrational mode specific to the zigzag nanotubes. Basically both the $L$ and $L''$ modes are associated with a rotative motion of the bonds out of the carbon plane. Bending due to the curvature of the tube has a different influence on stiffness against rotations of axial bonds (parallel to the tube axis), or circumferential bonds (rotated w.r.t. the tube axis), causing the different frequencies of the $L$ and $L''$ modes. With increasing tube diameter, this bending decreases and $L$ and $L''$ peaks merge into a single peak. Figure~\ref{fig:zigzag}(d) shows that the $L$ vibration mode corresponds to rotational motion of circumferential bonds. Figure~\ref{fig:zigzag}(e) shows that the $L''$ vibration mode corresponds to rotational motion of axial bonds. Since armchair tubes do not have bonds which are exactly parallel to the tube axis, they do not show a $L''$ peak in the VDOS. As the tube chirality changes from armchair tubes to zigzag tubes, we observe peak splitting ($L$ and $L''$) in the VDOS as discussed in the next section.

The $L$ modes shown in figure~\ref{fig:armchair}(d) and figure~\ref{fig:zigzag}(d) should be observable in IR and FTIR corresponding to the motion of homogeneous bond angle changes in the sample. We predict that the combination of modes from $L$ and $L'$ peaks might be Raman active but to make a definitive remark on it, one has to convert the current VDOS to the corresponding Raman spectra with the help of matrix elements. We believe that the modes corresponding to the $L$ peak should be as good as RBM to characterize the tubes. The main difference we see between $L$ and $L''$ modes is the change in the bond angles. In the RBM there are no changes in the bond angles since all the atoms collectively move inside or outside the carbon plane (breathing) but in the mode corresponding to the $L$ peak there are motions resulting into changes in the bond angles. Both $L$ and $L'$ peaks are acoustic and flexural and the effect of the diameter (curvature) is very significant on the VDOS.

We believe that for a small tube diameter there is electron-phonon coupling, as also explained in the paper by Maultzsch et al. \cite{maultzsch2005radial}. But since we are using a semiempirical potential to calculate the VDOS of the tube, we cannot explicitly verify the extent of this coupling. To effectively capture the effect of electron-phonon coupling in simulations, one should combine the semiempirical potential with ab-initio method or tight binding model.

These significant changes and unique signatures in the frequencies and intensities of low energy vibrational modes in armchair and zigzag nanotubes can be used to characterize the diameter of nanotubes with the help of the VDOS.

\subsection{Chiral nanotubes}
\begin{figure*}
\begin{subfigure}{0.58\textwidth}
(a)\\
\includegraphics[scale=1]{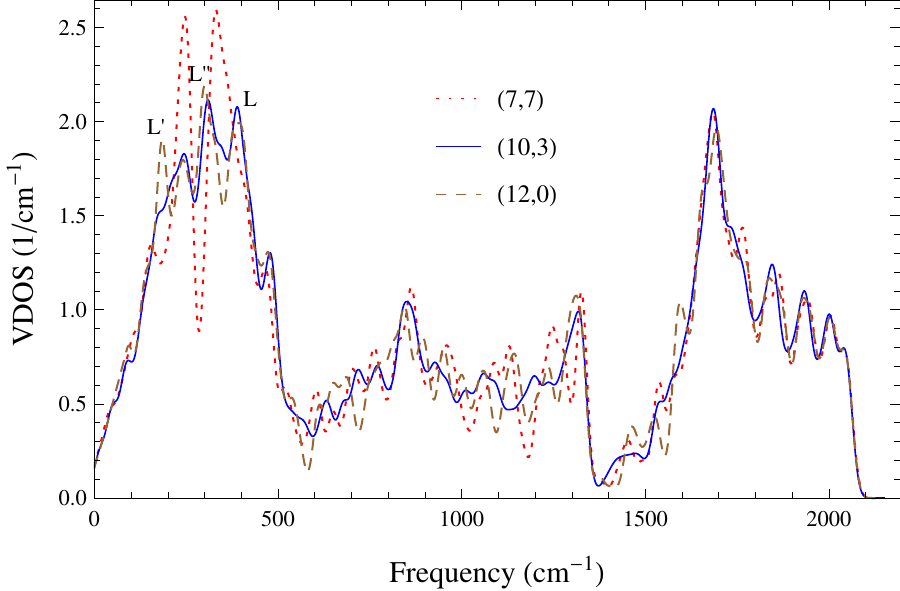}
\end{subfigure}
~
\begin{subfigure}{0.11\textwidth}
(b)\\
\includegraphics[scale=.05,trim=85cm 8cm 83cm 8cm,clip=true]{Fig-3b}\\
\centerline{$(7,7)$}
\end{subfigure}
~
\begin{subfigure}{0.09\textwidth}
(c)\\
\includegraphics[scale=.05,trim=88cm 8cm 88cm 8cm,clip=true]{Fig-3c}\\
\centerline{$(10,3)$}
\end{subfigure}
~
\begin{subfigure}{0.111\textwidth}
(d)\\
\includegraphics[scale=.05,trim=84cm 8cm 84cm 8cm,clip=true]{Fig-3d}\\
\centerline{$(12,0)$}
\end{subfigure}
\caption{Dependence of VDOS on chiral angle for different chiral tubes with almost the same diameter ($D\sim9.4$~\AA). (a) VDOS of three nanotubes: armchair $(7,7)$, chiral $(10,3)$ and zigzag $(12,0)$. (b-d) For illustration, the three chiralities of nanotubes used in (a).}
\label{fig:chiral}
\end{figure*}

In the previous sections, we show that for both armchair and zigzag nanotubes, the VDOS is highly dependent on tube diameter. Using tubes of different chiral angles with almost the same diameter, we study the effect of chiral angle on the VDOS. In figure~\ref{fig:chiral}, the VDOS of three such nanotubes is plotted: armchair $(7,7)$, chiral $(10,3)$ and zigzag $(12,0)$, with the diameters of $9.49\text{ \AA}$, $9.23\text{ \AA}$, and $9.39\text{ \AA}$ respectively.

Figure~\ref{fig:chiral} shows that there is a significant difference in the low frequency regime of the VDOS of tubes of equal diameter with different chiral angles. For the armchair $(7,7)$ tube, there are two prominent low energy $L$ and $L'$ peaks but as soon as we deviate from this chirality the $L$ peak starts to split into two, giving rise to the $L''$ peak. Comparing the $L$ and $L''$ peaks of the chiral $(10,3)$ and zigzag $(12,0)$ nanotubes to the $L$ peak of the armchair $(7,7)$ nanotube, we find that the frequency of the $L$ peak is $17\%$ higher for the $(10,3)$ nanotube and $19\%$ for the $(12,0)$ nanotube, and the frequency of the $L''$ peak is $6\%$ lower for the $(10,3)$ nanotube and $9\%$ lower for the $(12,0)$ nanotube. Compared to the $(7,7)$ nanotube, the frequency of the $L'$ peak is $8\%$ lower for the $(10,3)$ nanotube and $25\%$ lower for the $(12,0)$ nanotube. The effect of the chiral angle on the intensities of these soft modes is also very prominent. The intensity of the $L$ peak is $20\%$ lower for the $(10,3)$ nanotube and $23\%$ for the $(12,0)$ nanotube, and the intensity of the $L''$ peak is $18\%$ lower for the $(10,3)$ nanotube and $16\%$ for the $(12,0)$ nanotube. The intensity of the $L'$ peak is $28\%$ lower for the $(10,3)$ nanotube and $26\%$ for the $(12,0)$ nanotube. Therefore these significant changes in the frequencies and intensities of the $L$ and $L'$ peaks in the VDOS can be used to characterize the chirality of carbon nanotubes independent of their diameters.  

These small changes in diameter ($\sim 1\%$) can be distinguished by Raman spectroscopy by the RBM mode as discussed in the paper by Maultzsch et al.\cite{maultzsch2005radial}. Based on pattern recognition (frequency of RBM and energy of the optical transition); authors have suggested a way to distinguish various chiral nanotubes and assign the corresponding chiral indices to the tubes. Based on bulk fluorimetric intensities, Tsyboulski et al. \cite{tsyboulski2007structure} have explained in their paper that PL spectroscopy can be used to assign the chiral indices of nanotubes even for small changes in the diameter. These findings are consistent with other papers on experiments based on PL spectroscopy such as the work by Bachilo et al.\cite{bachilo2002structure} and Lefebvre et al.\cite{lefebvre2007polarized}.

\subsection{Stone--Wales defects}

\begin{figure*}
\begin{subfigure}{0.5\textwidth}
(a)\\
\includegraphics[scale=.9]{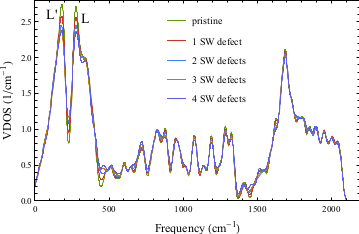}
\end{subfigure}
~
\begin{subfigure}{0.09\textwidth}
(b)\\
\includegraphics[scale=.035,trim=83cm 3cm 83cm 3cm,clip=true]{Fig-4b}
\end{subfigure}
~
\begin{subfigure}{0.09\textwidth}
(c)\\
\includegraphics[scale=.035,trim=82cm 3cm 81cm 3cm,clip=true]{Fig-4c}
\end{subfigure}
~
\begin{subfigure}{.25\textwidth}
(d)\\
\includegraphics[scale=.9]{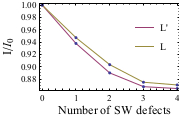}
\end{subfigure}

\begin{subfigure}{0.5\textwidth}
(e)\\
\includegraphics[scale=.9]{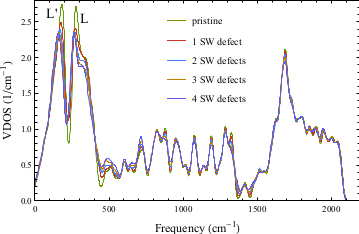}
\end{subfigure}
~
\begin{subfigure}{0.09\textwidth}
(f)\\
\includegraphics[scale=.035,trim=83cm 3cm 83cm 3cm,clip=true]{Fig-4f}
\end{subfigure}
~
\begin{subfigure}{0.09\textwidth}
(g)\\
\includegraphics[scale=.035,trim=82cm 3cm 81cm 3cm,clip=true]{Fig-4g}
\end{subfigure}
~
\begin{subfigure}{.25\textwidth}
(h)\\
\includegraphics[scale=.9]{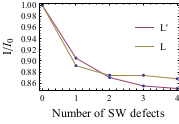}
\end{subfigure}

\caption{VDOS of armchair $(10,10)$ tube with one to four evenly spaced Stone--Wales defects. (a) VDOS for circumferential bond rotation (CBR) SW defect in armchair $(10,10)$ tube with different numbers of defects. (b) Front view of the tube with four SW defects. (c) Side view of the same tube. (d) Relative intensity of $L$ and $L'$ peaks as a function of number of CBR SW defects. (e) VDOS for axial bond rotation (ABR) SW defect in armchair $(10,10)$ tube with different numbers of defects. (f) Front view of the tube with four SW defects. (g) Side view of the same tube. (h) Relative intensity of $L$ and $L'$ peaks as a function of number of ABR SW defects.}
\label{fig:armchairsw}
\end{figure*}

We further study the effect of point defects such as Stone--Wales (SW) defects on the VDOS in two types of nanotubes: armchair $(10,10)$ and zigzag $(20,0)$, the diameters of the tubes are $13.56\text{ \AA}$ and $15.66\text{ \AA}$, respectively. An SW defect is formed by the rotation of a C--C bond by $90^{0}$ in a crystalline tube. In general, three types of bonds are found in a nanotube but in the case of armchair and zigzag chiralities, this reduces to two as two types of bonds are equal by symmetry. In armchair tubes, every carbon atom is connected to one circumferential and two equivalent axial bonds, thus the possible SW defects are obtained by either circumferential bond rotation (CBR) or axial bond rotation (ABR) \cite{dinadayalane2007stone}. We calculate the VDOS of a $(10,10)$ carbon nanotube with one to four evenly spaced SW defects of both kinds (CBR and ABR) as shown in figure~\ref{fig:armchairsw}.

We observe that in the $(10,10)$ armchair nanotube with CBR SW defect, the intensity of the $L$ and $L'$ peaks decreases systematically with an increasing number of SW defects compared to the pristine tube. The intensity of the $L$ peak decreases by $\sim 5\%$ for one SW defect to $\sim 13\%$ for four SW defects as shown in figure~\ref{fig:armchairsw}(d). Similarly the intensity of the $L'$ peak also decreases simultaneously by $\sim 6\%$ for one SW defect to  $\sim 14\%$ for four SW defects. For ABR SW defects in an armchair tube, the intensity of the $L$ peak decreases by $\sim 11\%$ for one SW defect and saturates at $13\%$ with a further increase in the number of defects as shown in figure~\ref{fig:armchairsw}(h). But at the same time, the intensity of the $L'$ peak decreases by $\sim 9\%$ for one SW defect to $\sim 14\%$ for three defects, and further decreases to $\sim 15\%$ for four SW defects. For CBR SW defects in armchair tube, the frequency of both peaks does not change with change in defect density. But in the case of ABR SW defects, there is a minor decrease (redshift) in the frequencies of the $L$ and $L'$ peaks. The frequency of the $L$ peak decreases by $\sim 2\%$ for one SW defect to $\sim 5\%$ for four SW defects. The frequency of the $L'$ peak decreases by $\sim 4\%$ for one SW defect to $\sim 14\%$ for four SW defects.

\begin{figure*}
\begin{subfigure}{0.5\textwidth}
(a)\\
\includegraphics[scale=.9]{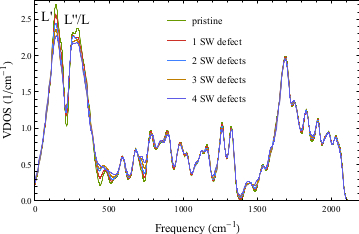}
\end{subfigure}
~
\begin{subfigure}{0.1\textwidth}
(b)\\
\includegraphics[scale=.03,trim=75cm 3cm 73cm 3cm,clip=true]{Fig-5b}
\end{subfigure}
~
\begin{subfigure}{0.09\textwidth}
(c)\\
\includegraphics[scale=.03,trim=78cm 3cm 78cm 3cm,clip=true]{Fig-5c}
\end{subfigure}
~
\begin{subfigure}{.24\textwidth}
(d)\\
\includegraphics[scale=.85]{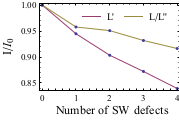}
\end{subfigure}

\begin{subfigure}{0.5\textwidth}
(e)\\
\includegraphics[scale=.9]{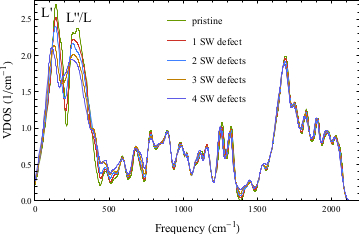}
\end{subfigure}
~
\begin{subfigure}{0.09\textwidth}
(f)\\
\includegraphics[scale=.03,trim=78cm 3cm 73cm 3cm,clip=true]{Fig-5f}
\end{subfigure}
~
\begin{subfigure}{0.1\textwidth}
(g)\\
\includegraphics[scale=.03,trim=75cm 3cm 74cm 3cm,clip=true]{Fig-5g}
\end{subfigure}
~
\begin{subfigure}{.24\textwidth}
(h)\\
\includegraphics[scale=.85]{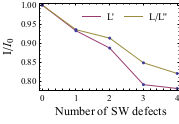}
\end{subfigure}

\caption{VDOS of zigzag $(20,0)$ tube with one to four evenly spaced Stone--Wales defects. (a) VDOS for circumferential bond rotation (CBR) SW defect in zigzag $(20,0)$ tube with different numbers of defects. (b) Front view of the tube with four SW defects. (c) Side view of the same tube. (d) Relative intensity of $L/L''$ and $L'$ peaks as a function of number of CBR SW defects. (e) VDOS for axial bond rotation (ABR) SW defect in zigzag $(20,0)$ tube with different numbers of defects. (f) Front view of the tube with four SW defects. (g) Side view of the same tube. (h) Relative intensity of $L/L''$ and $L'$ peaks as a function of number of ABR SW defects.}
\label{fig:zigzagsw}
\end{figure*}

In zigzag nanotubes, every carbon atom is connected to one axial bond and two circumferential bonds, leading to two possible configurations of SW defects (ABR and CBR). We calculate the VDOS as a function of SW defect density as shown in figure~\ref{fig:zigzagsw} for a $(20,0)$ zigzag tube with both ABR and CBR defect configurations. We show in figure~\ref{fig:zigzagsw}(a) and (e) that for this tube the $L$ and $L''$ peaks almost merge into a single peak. With increase in the defect density, the distinction between these two peaks becomes even more unclear therefore we plot only the maximum of the intensities of both $L$ and $L'$ peaks in the intensity plots as shown in figure~\ref{fig:zigzagsw}(d) and (h).

We observe that in the $(20,0)$ zigzag nanotube with CBR SW defect, the intensity of the $L/L''$ and $L'$ peaks decreases systematically with an increasing number of SW defects compared to the pristine tube. The intensity of $L/L''$ peak decreases by $\sim 4\%$ for one SW defect to $\sim 8\%$ for four SW defects as shown in figure~\ref{fig:zigzagsw}(d). The intensity of the $L'$ peak decreases more rapidly from $\sim 6\%$ for one SW defect to  $\sim 16\%$ for four SW defects. For ABR SW defects in zigzag tube, the intensity of the $L/L''$ peak decreases by $\sim 6\%$ for one SW defect to a significant $18\%$ for four SW defects as shown in figure~\ref{fig:zigzagsw}(h). But at the same time, the intensity of the $L'$ peak also decreases by $\sim 7\%$ for one SW defect to $\sim 21\%$ for three defects, and further decreases to $\sim 22\%$ for four SW defects. As in the case in armchair tubes, the frequencies of both peaks do not change with change in defect density in zigzag tubes with CBR SW defects. But in the case of ABR SW defects, there is a minor decrease in the frequency of the $L'$ peak. The frequency of the $L'$ peak decreases by $\sim 2\%$ for two SW defects to $\sim 19\%$ for four SW defects. This effect (redshift in the peak) is stronger in zigzag tubes compared to the armchair tubes suggesting that the axial bond rotation in zigzag nanotubes has the largest influence on the VDOS.

The study of effect of dopants on the VDOS of CNTs might be a nice future research topic. With our current semiempirical potential, we cannot simulate the effect of doping since all the parameters are fitted to carbon-only samples. The study of the effect of dopants in the tube will require new parameters in the current potential. We expect broadening of the peaks and a decreased intensity because of symmetry-breaking induced by the dopants. We also expect that there will be other peaks corresponding to the stretching and bending motions of the bonds made by dopants and carbon atoms. As in the case of SW defects in CNTs, we observe that symmetry-breaking introduced by the defects broadens the $L$ and $L'$ peaks because it reduces the degeneracy of the modes, resulting in modes with a small difference in frequency.

\section{CONCLUSION}
In conclusion, we have shown that the vibrational density of states can be used to characterize the structure of carbon nanotubes. With the help of a semiempirical potential, we calculated the vibrational spectrum of various carbon nanotubes and studied the characteristic low frequency $L$ and $L''$ modes. We show that the high frequency VDOS ($\geq 500\text{ cm}^{-1} $) of carbon nanotubes is almost identical to that of graphene but there are significant differences in the low energy VDOS. With our numerical simulations we have identified the Raman active radial breathing mode as the $L'$ peak in the VDOS of armchair nanotubes. There is redshift in both low frequency $L$ and $L''$ peaks as the diameter of the nanotube increases and the values of the frequencies converge to that of graphene in the infinite diameter limit.

Our results suggest that an $L$ peak in VDOS with a frequency over $400\text{ cm}^{-1}$, indicates a diameter~$\lesssim 8$ \AA. For nanotubes with a diameter~$\lesssim 15$ \AA, two sharp peaks close to each other in the low frequency range indicate an armchair chirality of the tube. When the two peaks are further apart and an extra third peak $L''$ peak appears in between, the tube can be characterize as chiral or zigzag nanotube. An $L$ peak with a frequency about twice that of the $L'$ peak, indicates a zigzag chirality.

Furthermore, we have studied the effect of point defects such as the Stone--Wales defect as a function of defect density on the vibrational spectrum of the nanotubes. Increasing the density of Stone--Wales defects leads to broader and less intense peaks, depending on the orientation of the defect with respect to the tube axis. For axial bond rotation there is redshift in the $L'$ peak for both armchair and zigzag tubes.

We hope that these results will stimulate further research into vibrational measurements via inelastic electron tunneling spectroscopy and neutron scattering spectroscopy on carbon nanotubes, 
or the vibrational properties of multi-walled carbon nanotubes \cite{pekker2011vibrational,cui2015heat} and other carbon nanostructures \cite{zhan2015carbon,cui2015enhancement,cui2016heat}. 
      
\section{Acknowledgements}
This project is funded by FOM-SHELL-CSER programme (12CSER049). This work is part of the research programme of the Foundation for Fundamental Research on Matter (FOM), which is part of the Netherlands Organisation for Scientific Research (NWO).

\bibliography{CNT}

\end{document}